\documentclass[12pt]{article}
\usepackage{amssymb,euscript,bbold}
\newcommand{\be}{\begin{equation}}
\newcommand{\ee}{\end{equation}}
\newcommand{\bea}{\begin{eqnarray}}
\newcommand{\eea}{\end{eqnarray}}
\newcommand{\ba}{\begin{array}}
\newcommand{\p}[1]{(\ref{#1})}
\newcommand{\ea}{\end{array}}

\def\bbox{{\,\lower0.9pt\vbox{\hrule \hbox{\vrule height 0.2 cm
\hskip 0.2 cm \vrule height 0.2 cm}\hrule}\,}}
\newcommand{\dsl}{\pa \kern-0.5em /}

\newcommand{\ep}{\epsilon}

\newcommand{\nn}{\nonumber \\}

\def\ds{\raise.15ex\hbox{/}\kern-.57em\partial}
\def\Ds{\,\raise.15ex\hbox{/}\mkern-13.5mu D}
\begin{document}

\makeatletter
\renewcommand{\theequation}{\thesection.\arabic{equation}}
\@addtoreset{equation}{section}
\makeatother

\baselineskip 18pt

%%%%%%%%%%%%%%%% title page %%%%%%%%%%%%%%%%%%%%%%%%%%%%%%%%%%%%

\begin{titlepage}
\vfill
\begin{flushright}
QMUL-PH-01-08\\
hep-th/0109039\\
\end{flushright}

\vfill

\begin{center}
\baselineskip=16pt
{\Large\bf M-Fivebranes Wrapped on} 
\\
{\Large\bf Supersymmetric Cycles II} 
\vskip 10.mm
{Jerome P. Gauntlett$^{1}$ and Nakwoo Kim$^{2}$\\}
\vskip 1cm
%\vfill
{\small\it
Department of Physics\\
Queen Mary, University of London\\
Mile End Rd, London E1 4NS, UK}\\
\vspace{6pt}
\end{center}
\vfill
\par
\begin{center}
{\bf ABSTRACT}
\end{center}
\begin{quote}
We construct D=11 supergravity solutions dual to
the twisted field theories arising when
M-theory fivebranes wrap supersymmetric cycles.
The cases considered are M-fivebranes wrapped on
(i) a complex Lagrangian four-cycle in a D=8 hyper-K\"ahler manifold
corresponding to a D=2 field theory with (2,1) supersymmetry
(ii) a product of two holomorphic two-cycles in a product
of two Calabi-Yau two-folds corresponding to a D=2 field
theory with (2,2) supersymmetry and (iii) a product
of a holomorphic two-cycle and a SLAG three-cycle in a product
of a Calabi-Yau two-fold and a Calabi-Yau three-fold corresponding
to a quantum mechanics with two supercharges.
In each case we construct BPS equations and find IR superconformal 
fixed points corresponding to new examples of AdS/CFT duality
arising from the twisted field theories. 
\vfill
\vskip 5mm
\hrule width 5.cm
\vskip 5mm
%{\small
\noindent $^1$ E-mail: j.p.gauntlett@qmw.ac.uk \\
\noindent $^2$ E-mail: n.kim@qmw.ac.uk \\
%}
\end{quote}
\end{titlepage}
%%%%%%%%%%%%%%%%%%%%%%%%%%%%%%%%%%%%%%%%
\setcounter{equation}{0}

\section{Introduction}

An interesting way to generalise the AdS/CFT 
correspondence \cite{mal} is to construct supergravity duals of the 
twisted field theories arising \cite{bvs} when branes wrap supersymmetric 
cycles. By exploiting the observation that the 
supergravity solutions can be first constructed in an appropriate 
gauged supergravity and then uplifted to D=10 or D=11, such
solutions were presented in \cite{malnun}.
The examples considered in \cite{malnun} involve
M-fivebranes and D-3-branes wrapping two-cycles in 
Calabi-Yau two- or three-folds. 
The solutions describe a flow from a UV region, corresponding to the 
D=6 or D=4 twisted field theory on the brane wrapped on the cycle, 
to an IR region corresponding to a D=4 or D=2 dimensional field
theory, where the energy scale is set by the inverse size of the cycle.
In several cases AdS fixed points were found in the IR corresponding to new 
AdS/CFT examples.

In subsequent work, D=11 supergravity solutions corresponding 
to M-fivebranes wrapped on associative three-folds in manifolds of
$G_2$ holonomy were constructed in \cite{agk} and many other 
examples were considered in \cite{gkw}. 
Other cases involving D3-branes and M2-branes were considered in
\cite{no} and \cite{gkpw}, respectively. In addition to these
examples involving conformal branes, analogous supergravity solutions 
for other wrapped branes have been studied in 
\cite{malnuntwo,agk,npst,en,sch,mnast,hern,gkmw,bcz,goma}.

Here we would like to report on three outstanding cases involving
M-fivebranes. As in \cite{gkw}, the solutions are constructed in 
maximal D=7 gauged supergravity \cite{vn} 
and then uplifted to D=11 using the results of \cite{vntwo,vnthree}. 
Indeed, we shall employ exactly the same techniques as \cite{gkw} 
and we refer the reader to this paper for further background and details on 
notation and conventions. 
The first case we consider, which is probably the most
interesting, is M-fivebranes wrapping a 
``complex Lagrangian'' four-cycle in a D=8 hyper-K\"ahler manifolds.
These are four-cycles that are complex (K\"ahler) with respect to one of 
the three complex structures and special Lagrangian
with respect to another, which together imply that they are also
special Lagrangian with respect to the third complex structure
\cite{dadok} (for a discussion in the physics literature, see \cite{jmf}). 
A concrete example of such a supersymmetric four-cycle is 
$CP^2$ in the hyper-K\"ahler Calabi metric on $T^*(CP^2)$.
At low-energies the wrapped fivebrane  
gives rise to a D=2 field theory preserving 
(2,1) supersymmetry, as we shall show. 
In the supergravity solutions we construct, the metric on the 
four-cycle $\Sigma_4$  is taken to be K\"ahler with constant holomorphic 
sectional curvature. In other words, $CP^2$ for positive curvature;
the Bergmann metric on a unit open ball $D^2$ in $C^2$ for negative curvature;
and flat space for zero curvature (see p. 170 of \cite{kn}).
As in previous solutions we can also take a quotient of these spaces
by a discrete group of isometries and in 
particular we can obtain compact manifolds
with negative curvature. We construct the BPS equations and demonstrate
an $AdS_3\times \Sigma_4$ IR fixed point when $\Sigma_4$ has negative
curvature and we determine the central charge of the fixed point.
We show that the numerical analyses of the BPS equations is 
essentially included in \cite{gkw}.

The second case to be considered is M-fivebranes wrapping a product of
two two-cycles, $\Sigma_1\times \Sigma_2$, with each $\Sigma_i$
a K\"ahler (holomorphic) two-cycle in a Calabi-Yau two-fold.
At low-energies this gives rise to a 
D=2 field theory with (2,2) supersymmetry.
We find BPS equations when each $\Sigma_i$ has constant curvature and
show that there is an IR $AdS_3\times \Sigma_1\times \Sigma_2$
fixed point in the special case that the four-cycle is Einstein 
with negative curvature. We again determine the central charge of 
the fixed point. 

The final case we will examine is M-fivebranes wrapping a product of
a three-cycle with a two-cycle, $\Sigma_1\times \Sigma_2$, with  
$\Sigma_1$ a SLAG three-cycle in  a Calabi-Yau three-fold and
$\Sigma_2$ a K\"ahler two-cycle in  a Calabi-Yau two-fold.
At low-energies this gives rise to a quantum mechanics with 2 supercharges.
We find BPS equations when each $\Sigma_i$ has constant curvature and
show that there is an IR $AdS_2\times \Sigma_1\times \Sigma_2$ 
fixed point in the special case that the five-cycle
is Einstein with negative curvature.

\section{Four-Cycles in D=8 Hyper-K\"ahler Manifolds}

In this section we consider fivebranes 
wrapping supersymmetric complex Lagrangian four-cycles 
in D=8 hyper-K\"ahler manifolds. Before turning to the
construction of the supergravity solutions let us begin
by describing the supersymmetry preserved by a probe fivebrane
wrapping such a cycle. It will then be straightforward to
impose the appropriate supersymmetry projections in the gauged
supergravity context. Consider a D=11 background to be of
the form $R^{1,2}\times M$ where $M$ is a hyper-K\"ahler eight-manifold.
It will be convenient to introduce an orthonormal frame 
$e^a$, $a=1,\dots,8$, with hyper-K\"ahler structure given by
\bea\label{hkstruct}
J^1&=&e^1\wedge e^2-e^3\wedge e^4-e^5\wedge e^6 +e^7\wedge e^8\nn
J^2&=&e^1\wedge e^5+e^2\wedge e^6+e^3\wedge e^7 +e^4\wedge e^8\nn
J^3&=&e^1\wedge e^6-e^2\wedge e^5-e^3\wedge e^8 +e^4\wedge e^7=J^1J^2
\eea
Noting that $(1/2)J^1\wedge J^1$, 
$-Re\Omega^{J^2}\equiv -Re(e^1+ie^5)(e^2+ie^6)(e^3+ie^7)(e^4+ie^8)$
and $-Re\Omega^{J^3}\equiv -Re(e^1+ie^6)(e^2-ie^5)(e^3-ie^8)(e^4+ie^7)$
can be expressed as $-e^{1234}+\dots$ we conclude that 
a four-cycle whose volume form is the pull-back of $-e^{1234}$ 
is complex with respect to $J^1$ and 
special Lagrangian with respect to $J^2,J^3$. 
If we wrap a fivebrane probe on this four-cycle
the D=11 supersymmetry preserved satisfies
\bea\label{elprojs}
\Gamma^{091234}\epsilon=\epsilon\nn
\Gamma^{12}\epsilon=-\Gamma^{56}\epsilon\nn
\Gamma^{34}\epsilon=-\Gamma^{78}\epsilon\nn
(\Gamma^{14}+\Gamma^{23}+\Gamma^{58}+\Gamma^{67})\epsilon&=
&0
\eea
where the last three conditions are the projections imposed
on the parallel spinors of the hyper-K\"ahler manifold (see, for example,
\cite{cglp}), and the first is due to the wrapped fivebrane. These 
conditions preserve 3/32 supersymmetry or more precisely
(2,1) supersymmetry, where the chirality refers to the D=2 field 
theory living on the unwrapped directions of the fivebrane. To see
this, first note that the first three projections preserve 1/8 supersymmetry.
Next note that the last condition can be replaced with
$(1-S_1-S_2-S_3)\epsilon =\epsilon$ where 
$S_1\equiv \Gamma^{1234}$, $S_2\equiv \Gamma^{1458}$ and 
$S_3\equiv\Gamma^{2358}$. The $S_i$ all have vanishing trace,
commute with the other projectors, square to unity and
satisfy $S_1S_2S_3=-1$.
Working on the subspace where the 
other conditions in \p{elprojs} are satisfied 
we can choose a basis where 
$S_1=diag(1,1,-1,-1)$, $S_2=diag(1,-1,1,-1)$ and $S_3=diag(-1,1,1,-1)$.
It is then easy to see that the last condition preserves 3/4 of the 
supersymmetry corresponding to the spinors $\ep_1=(1,0,0,0)$,
$\ep_1=(0,1,0,0)$, $\ep_1=(0,0,1,0)$. In addition, from the first condition
in \p{elprojs} we note that the D=2 chirality is specified by the action of 
$\Gamma^{1234}$ and we see that $\ep_1$ and $\ep_2$ have positive
helicity and $\ep_3$ has negative helicity giving rise to 
(2,1) supersymmetry as claimed.  

It is interesting to observe that the spinors $\ep_1$ and 
$\ep_2$ are annihilated by $S_2+S_3$ and hence 
for these spinors the last condition is simply $(1-S_1)\epsilon=0$.
Comparing with, e.g. \cite{glw}, one now sees that this projection 
along with the first three in \p{elprojs} are precisely those for a 
K\"ahler four cycle corresponding to $J^1$.
Similarly one finds that $\ep_1$ and $\ep_3$ are associated with
the projections for a SLAG four-cycle with respect to the complex
structure $J^2$ and $\ep_2$ and $\ep_3$ are associated with
the projections for a SLAG four-cycle with respect to
$J^3$.

Having finished this explicit discussion of the supersymmetry
projections for the fivebrane probe we are ready to start with 
the construction of the corresponding gauged supergravity solutions.
As noted, we shall first construct the solutions
in maximal D=7 gauged supergravity \cite{vn} and we refer the 
reader to \cite{gkw} for more details on notation. 
The ansatz for the D=7
metric is given by
\be
ds^2_7=e^{2f}(-dt^2+dx^2+dr^2)+e^{2g}d\bar s^2 
\ee
where $t,x$ are coordinates of the unwrapped part of the fivebrane 
worldvolume,
$d\bar s^2$ is the metric on the four-cycle $\Sigma_4$
that the fivebrane wraps and $f,g$ are functions of $r$ only. 

The ansatz for the $SO(5)$ supergravity gauge
fields are directly determined by the ``twisting'' arising 
when an M-fivebrane probe wraps a supersymmetric cycle.
This twisting is simply a consequence of the structure 
of the normal bundle of the supersymmetric cycle\cite{bvs}.
It entails an identification of the structure group of the cycle 
with a subgroup of the $SO(5)$ R-symmetry and is required in 
order to preserve supersymmetry.
We can thus determine the $SO(5)$ supergravity
gauge field ansatz by consideration of the supersymmetry 
preserved by the M-fivebrane wrapping the four-cycle. 
In the language of  gauged supergravity the appropriate 
supersymmetry projections discussed above are given by
\bea
\label{hk}
\gamma^r\epsilon&=&\epsilon\nn
(1-\gamma^{12}\Gamma^{12})\epsilon&=&0\nn
(1-\gamma^{34}\Gamma^{34})\epsilon&=&0\nn
(\gamma^{14}+\gamma^{23}+\Gamma^{14}+\Gamma^{23})\epsilon&=
&0
\eea
where $\gamma^\mu$ and $\Gamma^m$ are $SO(1,6)$ and $SO(5)$ gamma-matrices,
respectively, the indices refer to an obvious orthonormal frame and 
the directions $1,2,3,4$ correspond to those of the cycle. By repeating
a similar analysis to that above we conclude that these projections
preserve (2,1) supersymmetry, where the chirality refers to the D=2 field 
theory living on the unwrapped directions of the fivebrane specified. 

The ``twisting condition'' required by supersymmetry is given by \cite{gkw}:
\be\label{twist}
(\bar\omega_{ab}\gamma^{ab}+2mB_{mn}\Gamma^{mn})\ep =0
\ee
where  $\bar\omega$ is the spin connection of 
$\Sigma_4$ with $a,b=1,\dots,4$ tangent space indices, 
and $B$ is the SO(5) gauge-field with $m,n=1,\dots,5$.
Upon imposing the projections we see that this condition is satisfied 
if we  demand that the metric on the cycle is K\"ahler 
(i.e. impose $\bar\omega_{31}=\bar\omega_{24}$ and 
$\bar\omega_{23}=\bar\omega_{14}$ corresponding to the K\"ahler form
with non-vanishing entries given by
$J_{12}=-J_{34}=1$) and in addition we demand that the only non-vanishing
gauge fields are in a $U(2)$ subgroup of $SO(5)$ and identified with
the spin connection via $\bar\omega=2m B$.
In other words, we see that when a fivebrane wraps a complex 
Lagrangian four-cycle 
in a D=8 hyper-K\"ahler manifold, the
appropriate twisted field theory is obtained by identifying the $U(2)$ 
spin connection of the cycle with a corresponding $U(2)$ 
subgroup of the $SO(5)$ R-symmetry.

As in \cite{gkw}, with the type of ansatz we consider,
supersymmetry demands that the K\"ahler four-cycle is 
Einstein, and we take $\bar R_{ab}=l\bar g_{ab}$ with $l=\pm 1,0$.
To ensure all equations of motion are satisfied we demand
in addition that it
has constant holomorphic sectional curvature\footnote{
Another way to satisfy the equations of motion and preserve (2,2)
supersymmetry, is to take the four-cycle to be a product of 
two constant curvature two-metrics. This will be discussed in the next 
section.}. Equivalently, we demand that the Riemann tensor of the
four-cycle can be expressed as
\be
\bar R_{ab}^{\,\,\,\,\, cd}={l\over 3}\left[J_{ab} J^{cd}+\delta^{cd}_{ab}+
J_a^{\,\,\, [c}J_b^{\,\,\, d]}\right]
\ee
For $l=1$ we have $CP^2$, for $l=0$ flat space and 
for $l=-1$ the Bergmann metric on the unit ball $D^2$ in $C^2$, 
or a quotient of these spaces by a discrete group of isometries.

We truncate the 15 scalar fields of maximal gauge supergravity 
to a single scalar field $\lambda(r)$. 
As for the other cases of M-fivebranes wrapping four-cycles in eight 
dimensions considered in \cite{gkw}, and consistent
with the twisting just discussed, we take
\be\label{slagfourscalars}
{\Pi_A}^i=(e^{\lambda},e^{\lambda},e^{\lambda},
         e^{\lambda},e^{-4\lambda})\, .
\ee
Only one of the five three-forms, $S_5$, is non-zero and is given by
\be\label{S4def}
S_5=-{\; e^{-8\lambda-4g+3f}\over 3{\sqrt 3} m^4} \, 
dt\wedge dx\wedge dr\, .
\ee
%\bea
%c=\epsilon^{a_1a_2a_3a_4}
%\epsilon^{b_1b_2b_3b_4}\bar R_{a_1a_2b_1b_2}\bar R_{a_3a_4b_3b_4}=64/3
%\eea

By setting the supersymmetry variations of the D=7 fermions to zero we
find that the resulting BPS equations are given by
\bea\label{slagfour}
e^{-f}f'&=& - {m\over 10}\left[4e^{-2\lambda}+e^{8\lambda}\right]
+{l \over 5m }{e^{2 \lambda-2g}}
-{l^2\over 5 m^3}{e^{-4\lambda-4g}}\nn
e^{-f}{g'} &=& - {m\over 10}\left[4e^{-2\lambda}+e^{8 \lambda}\right]
- {3l \over 10m }{e^{2 \lambda-2g}}
+{2l^2\over 15 m^3}{e^{-4\lambda-4g}}\nn
e^{-f}\lambda' &= & {m\over 5}\left[ 
e^{8\lambda} - e^{-2\lambda}\right] +{l\over 10m } {e^{2\lambda-2g}}
+{l^2\over 15 m^3}{e^{-4\lambda-4g}}\, .
\eea
Any solution of these BPS equations, and others presented in
the next sections, also solves the full equations of motion.
We do not have a general solution to these BPS equations. However,
the numerical analyses carried out for the BPS equations for
other four-cycles in \cite{gkw}
is applicable here (set $\alpha=l/m$,$\beta=2/3m^3$ in equation (6.14) of 
\cite{gkw}). In particular figures 5 and 6 of \cite{gkw} 
illustrate the 
corresponding behaviour of the flows from the UV to the IR for $l=\pm 1$. 

Using the results of \cite{vn,vntwo} we can uplift solutions to the BPS
equations to give supersymmetric solutions to $D=11$ supergravity. 
The metric is given by 
\be\label{hyperfourup}
ds^2_{11}=\Delta^{-{2\over 5}}ds^2_{7} +{1\over m^2}\Delta^{4\over 5}
\left[e^{2\lambda}DY^aDY^a+e^{-8\lambda}dY^5dY^5\right]
\ee
where 
\bea\label{hyperfouruptwo}
DY^a&=&dY^a+\bar\omega^{ab}Y^b\nn
\Delta^{-{6\over 5}}&=&e^{-2\lambda}Y^aY^a+e^{8\lambda}Y^5Y^5
\eea
and $(Y^a,Y^5)$ are constrained coordinates
on the four-sphere satisfying $Y^aY^a+Y^iY^i=1$. 
The expression for the four-form
can easily be read from the formulae in \cite{vn,vntwo} and we will
not bother to write it explicitly here.

If we take the four-cycle to have constant negative holomorphic
sectional curvature, $l=-1$, we
find that the BPS equations admit an $AdS_3\times \Sigma_4$ 
solution with:
\bea\label{sfscft}
e^{10\lambda}&=&{6\over 5}\nn
e^{2g}&=&{e^{-6\lambda}\over m^2}\nn
e^f&=&{e^{2\lambda}\over m}{1\over r}\, .
\eea
The central charge of the corresponding D=2 superconformal field theory can
be obtained from the radius of $AdS_3$. Repeating the arguments
in  \cite{gkw}, we find, setting $m=2$,
\be
c={8N^3\over \pi^2}{5\over 192} Vol(\bar\Sigma)\, .
\ee

It is worth mentioning how this example interconnects
with the more general class of solutions corresponding to fivebranes
wrapping K\"ahler and SLAG 
four-cycles in Calabi-Yau four-folds
discussed in \cite{gkw}. In that paper, it was shown
that if the four-cycle is K\"ahler-Einstein with the $U(1)\subset
U(2)$ part of the spin connection identified with the corresponding 
$U(1)$ of $U(2)\subset SO(5)$ then there are BPS equations
preserving $(2,0)$ supersymmetry. The analysis in \cite{agk}
only covered the case when the rest of the $U(2)\subset SO(5)$ 
gauge fields vanished. Here we have shown that if they are switched
on, for the special case when the full $U(2)$ gauge fields 
are identified with the $U(2)$ spin connection, and
in addition the K\"ahler-Einstein metric is taken to have constant
holomorphic sectional curvature, we get BPS equations preserving 
(2,1) supersymmetry. The supersymmetric four-cycles we are considering
in a hyper-K\"ahler manifold are also SLAG four-cycles
(with respect to a different complex structure).
The reason that the BPS equations presented here were not included in 
the SLAG four-cycles considered in \cite{gkw} is that it was assumed
there that the structure group of the four-cycle was in fact $SO(4)$
and not a proper subgroup of it. 

Another specialisation, to be discussed in
the next section, is when the supersymmetric
four-cycle is taken to be a product of two constant curvature
metrics. In this case the
structure group of the spin connection of the four-cycle
is $U(1)\times U(1)$ and this is identified with 
a corresponding $U(1)\times U(1)\subset SO(5)$. With these restrictions,
(2,2) supersymmetry is preserved. Again this can be viewed as
a special case of a SLAG or K\"ahler four-cycles with supersymmetry 
enhanced to (2,2).

\section{A product of two K\"ahler two-cycles}

Consider M-fivebranes wrapping a four-cycle consisting of a
product of two-cycles, $\Sigma_1\times \Sigma_2$, with each
$\Sigma_i$ a K\"ahler two-cycle in a Calabi-Yau two-fold. 
This is another example of an M-fivebrane wrapped on a four-cycle
in eight dimensions, but the product structure allows us to 
consider a slightly more general ansatz for the metric than in 
the previous example. Specifically, we now take
\be
ds^2_7=e^{2f}(-dt^2+dx^2+dr^2)+e^{2g_1}d\bar s^2_1 +e^{2g_2}d\bar s^2_2
\ee
with $d\bar s^2_i$ two-metrics on each of the two-cycles, and $f,g_1,g_2$
functions of $r$.
Combined with the twisting to be discussed, supersymmetry
forces these metrics to have constant curvature with
$\bar R_{ab}=l_i\bar g_{ab}$ and $l_i=\pm 1,0$. Each two-cycle must
be  $S^2$ for positive, flat space for zero curvature, $H^2$ 
for negative curvature, and again we can take quotients of these spaces
by discrete isometry subgroups. 
Note that this ansatz allows
for the four-cycle to be non-Einstein in general.

The appropriate supersymmetry projections are now given by
\bea
\label{tt}
\gamma^r\epsilon&=&\epsilon\nn
(1-\gamma^{12}\Gamma^{12})\epsilon&=&0\nn
(1-\gamma^{34}\Gamma^{34})\epsilon&=&0\, .
\eea
These preserve 1/8 of the supersymmetry, or more precisely,
(2,2) supersymmetry from the point of view of the unwrapped
D=2 part of the M-fivebrane world-volume. These projections
give rise to the appropriate ansatz for the $SO(5)$ gauge-fields
via the twisting condition \p{twist}. We split $SO(5)\to U(1)\times U(1)$
and identify each $U(1)$ with a $U(1)$ factor of the $U(1)\times U(1)$
structure group of the four-cycle. In other words we set
$\bar\omega_{12}=2m B_{12}$, $\bar\omega_{34}=2m B_{34}$  and all
other gauge fields vanishing. Clearly this is just two copies of the twisting
of a holomorphic 
two-cycle inside a Calabi-Yau two-fold considered in \cite{malnun}.

We choose a two-scalar ansatz consistent with $U(1)\times U(1)$ symmetry
via
\be\label{twotwoscalar}
{\Pi_A}^i=(e^{\lambda_1},e^{\lambda_1},e^{\lambda_2},
         e^{\lambda_2},e^{-2\lambda_1-2\lambda_2})
\ee
with $\lambda_i$ functions of $r$.
Only one of the five three-forms, $S_5$, is non-vanishing and is given by
\be\label{Sdef}
S_5=-{l_1 l_2e^{-4\lambda_1-4\lambda_2-2g_1-2g_2+3f}\over 2 {\sqrt 3} m^4} \, 
dt\wedge dx
\wedge dr\, .
\ee

Setting the supersymmetry variations of the fermions to zero
we obtain the following BPS equations
\bea
e^{-f}f'&=& - {m\over 10}\left[2e^{-2\lambda_1}+2e^{-2\lambda_2}+
e^{4\lambda_1+4\lambda_2}\right]
+{1 \over 10m }\left[l_1e^{2 \lambda_1-2g_1}+l_2e^{2 \lambda_2-2g_2}\right]
-{3l_1l_2\over 10 m^3}X\nn
e^{-f}{g_1'} &=& - {m\over 10}\left[2e^{-2\lambda_1}+2e^{-2\lambda_2}+
e^{4 \lambda_1+4\lambda_2}\right]
- {1 \over 10m }\left[4l_1e^{2 \lambda_1-2g_1}-l_2e^{2 \lambda_2-2g_2}\right]
+{l_1l_2\over 5 m^3}X\nn
e^{-f}{g_2'} &=& - {m\over 10}\left[2e^{-2\lambda_1}+2e^{-2\lambda_2}+
e^{4 \lambda_1+4\lambda_2}\right]
- {1 \over 10m }\left[4l_2e^{2 \lambda_2-2g_2}-l_1e^{2 \lambda_1-2g_1}\right]
+{l_1l_2\over 5 m^3}X\nn
e^{-f}\lambda_1' &= & {m\over 5}\left[ 
e^{4\lambda_1+4\lambda_2} - 3e^{-2\lambda_1}+2e^{-2\lambda_2}\right] 
+{1\over 10m } \left[{3l_1e^{2\lambda_1-2g_1}}-2l_2e^{2\lambda_2-2g_2}\right]
+{l_1l_2\over 10 m^3}X\nn
e^{-f}\lambda_2' &= & {m\over 5}\left[ 
e^{4\lambda_1+4\lambda_2} - 3e^{-2\lambda_2}+2e^{-2\lambda_1}\right] 
+{1\over 10m } \left[{3l_2e^{2\lambda_2-2g_2}}-2l_1e^{2\lambda_1-2g_1}\right]
+{l_1l_2\over 10 m^3}X\nn
&&
\eea
where $X\equiv e^{-2\lambda_1-2\lambda_2-2g_1-2g_2}$. 

The metric of the corresponding D=11 supergravity solution now has
the form
\be
ds^2_{11}=\Delta^{-{2\over 5}}ds^2_{7} +{1\over m^2}\Delta^{4\over 5}
\left[e^{2\lambda_1}DY^aDY^a+e^{2\lambda_2}DY^\alpha DY^\alpha
+e^{-4\lambda_1-4\lambda_2}dY^5dY^5\right]
\ee
with $a,b=1,2$, $\alpha,\beta =3,4$ and
\bea
DY^a&=&dY^a+\bar\omega^{ab}Y^b\nn
DY^\alpha&=&dY^a+\bar\omega^{\alpha\beta}Y^\beta\nn
\Delta^{-{6\over 5}}&=&e^{-2\lambda_1}Y^a Y^a+e^{-2\lambda_2}Y^\alpha Y^\alpha+
e^{4\lambda_1+4\lambda_2}Y^5Y^5
\eea
and $(Y^a,Y^\alpha,Y^5)$ are again constrained coordinates on the four-sphere. 
The expression for the four-form can be read off from the
formulae in \cite{vn,vntwo}.

It is straightforward to show
that the only $AdS_3$ fixed point of the BPS equations has $l_1=l_2=-1$ and
\bea
\lambda_i&=&0\nn
e^{2g_i}&=&{1\over m^2}\nn
e^f&=&{1\over m}{1\over r}\, .
\eea
Note that the four-cycle is $H_2\times H_2$ or a quotient thereof and
is Einstein.
In addition the warp factor in D=11 is trivial, so the
D=11 metric is simply a 
twisted product of $AdS_3\times H_2\times H_2\times S^4$.
The central charge of the D=2 superconformal field theory can
be obtained from the radius of $AdS_3$. Repeating the arguments
in  \cite{gkw} we find, upon setting $m=2$,
\be
c={8N^3\over \pi^2}{1\over 32} Vol(\bar\Sigma)\, . 
\ee

We shall not attempt a numerical analysis of the general BPS equations here.
However, if we restrict to the case $g_1=g_2$, $\lambda_1=\lambda_2$ 
and $l_1=l_2$, the numerical
analyses carried out for the BPS equations for other four-cycles in \cite{gkw}
is applicable here (set $\alpha=l_1/m$,$\beta=1/m^3$ in equation (6.14) of 
\cite{gkw}). In particular figures 5 and 6 of \cite{gkw} illustrate the 
corresponding behaviour of the flows from the UV to the IR.

\section{A product of a SLAG three-cycle with a K\"ahler two-cycle}

Our final example concerns M-fivebranes wrapping a 
five-cycle consisting of a product of a SLAG three-cycle in a
Calabi-Yau three-fold with a K\"ahler two-cycle in a 
Calabi-Yau two-fold. The metric ansatz is taken to be
\be
ds^2_7=e^{2f}(-dt^2+dr^2)+e^{2g_1}d\bar s^2_1 +e^{2g_2}d\bar s^2_2
\ee
with $d\bar s^2_1$ a three-metric of constant curvature with
$\bar R_{ab}=l_1g_{ab}$ and $d\bar s^2_2$ a two-metric of constant 
curvature with $\bar R_{\alpha\beta}=l_2g_{\alpha\beta}$ with $l_i=\pm 1,0$.
As usual these restrictions on the metrics arise from supersymmetry and
the equations of motion.

The supersymmetry projections for this case are given by
\bea
\label{tth}
\gamma^r\epsilon&=&\epsilon\nn
\gamma^{ab}\ep&=&-\Gamma^{ab}\epsilon\nn
\gamma^{\alpha\beta}\ep&=&-\Gamma^{\alpha\beta}\epsilon
\eea
where $a,b=1,2,3$, $\alpha,\beta=4,5$
and preserve 1/16 of the supersymmetry i.e. the low-energy
effective quantum mechanics arising from the wrapped fivebrane
has two supercharges. These projections
give rise to the following twisting. The spin connection has
structure group $SO(3)\times SO(2)$ and we identify this with
a corresponding subgroup of the $SO(5)$ R-symmetry. Concretely
we set $\bar\omega_{ab}=2m B_{ab}$,
$\bar\omega_{\alpha\beta}=2m B_{\alpha\beta}$
and set all other components of the gauge-fields to zero. 
This twisting is just
the combination of the twisting associated with the SLAG three-cycle 
discussed in \cite{gkw} and the K\"ahler two-cycle discussed in \cite{malnun}.

An ansatz for the scalar fields preserving the $SO(3)\times SO(2)$
symmetry is given by
\be
{\Pi_A}^i=(e^{2\lambda},e^{2\lambda},e^{2\lambda},e^{-3\lambda},
e^{-3\lambda})\, .
\ee
 Three of the five three-forms are active and we have
\be
S_m=-{l_1l_2e^{-2g_1-2g_2+4\lambda+2f} \over 4{\sqrt 3}m^4}\, 
dt\wedge dr\wedge e^m\, ,
\ee
for $m=1,2,3$.
With this ansatz the resulting BPS equations are given by 
\bea
e^{-f}f'&=& - {m\over 10}\left[3e^{-4\lambda}+2e^{6\lambda}\right]
+{3l_1 \over 20m }{e^{4 \lambda-2g_1}}
+{l_2 \over 10m }{e^{-6 \lambda-2g_2}}
-{9l_1l_2\over 20 m^3}{e^{2\lambda-2g_1-2g_2}}\nn
e^{-f}{g_1'} &=& - {m\over 10}\left[3e^{-4\lambda}+2e^{6 \lambda}\right]
- {7l_1 \over 20m }{e^{4 \lambda-2g_1}}
+ {l_2 \over 10m }{e^{-6 \lambda-2g_2}}
+{l_1l_2\over 20 m^3}{e^{2\lambda-2g_1-2g_2}}\nn
e^{-f}{g_2'} &=& - {m\over 10}\left[3e^{-4\lambda}+2e^{6 \lambda}\right]
+ {3l_1 \over 20m }{e^{4 \lambda-2g_1}}
- {2l_2 \over 5m }{e^{-6 \lambda-2g_2}}
+{3l_1l_2\over 10 m^3}{e^{2\lambda-2g_1-2g_2}}\nn
e^{-f}\lambda' &= & {m\over 5}\left[ 
e^{6\lambda} - e^{-4\lambda}\right] 
+{l_1\over 10m } {e^{4\lambda-2g_1}}
-{l_2\over 10m } {e^{-6\lambda-2g_2}}
-{l_1l_2\over 20 m^3}{e^{2\lambda-2g_1-2g_2}}\, .
\eea
The corresponding D=11 metric is given by
\be
ds^2_{11}=\Delta^{-{2\over 5}}ds^2_{7} +{1\over m^2}\Delta^{4\over 5}
\left[e^{4\lambda}DY^aDY^a+e^{-6\lambda}DY^\alpha DY^\alpha\right]
\ee
where
\bea
DY^a&=&dY^a+\bar\omega^{ab}Y^b\nn
DY^\alpha&=&dY^\alpha+\bar\omega^{\alpha\beta}Y^\beta\nn
\Delta^{-{6\over 5}}&=&e^{-4\lambda}Y^aY^a+e^{6\lambda}Y^\alpha Y^\alpha
\eea
and $(Y^a,Y^\alpha)$ are constrained coordinates
on the four-sphere. The expression for the four-form
can again be easily found using the formulae in \cite{vn,vntwo}.        

We find that there is only an $AdS_2$ fixed point 
when $g_1=g_2$ and $l_1=l_2=-1$.
\bea
e^{10\lambda}&=&2\nn
e^{2g_i}&=&{e^{8\lambda}\over 2 m^2}\nn
e^f&=&{e^{4\lambda}\over 2m}{1\over r}\, .
\eea
In particular the product manifold on the 5-cycle is $H^2\times H^3$ or
a quotient thereof and is Einstein.

\section{Discussion}
We have discussed new D=11 supergravity solutions describing
M-fivebranes wrapped on supersymmetric cycles. We have demonstrated
new AdS fixed points corresponding to the CFTs arising from
the twisted M-fivebrane theory in the IR. 
It would be interesting if we can compare our supergravity solutions
directly with the field theory arising on the M-fivebrane. 
For example it may be possible
to check the central charges of the CFT fixed points that we derived
from the supergravity point of view.

By simply 
enumerating cases it would seem that the solutions discussed 
here and those in \cite{malnun,gkw} cover\footnote{Note that 
solutions for M-fivebranes fully wrapping a Calabi-Yau two-fold were
discussed in \cite{ggpt}.} all ways in which static M-fivebranes 
can wrap supersymmetric cycles in a non-trivial Riemannian 
manifold with parallel spinors. Of course our ansatz can be generalised in a
number of ways and it would be interesting if more solutions can be found
in closed form. It is worth noting that
there are additional configurations of branes intersecting at angles
in flat space that preserve supersymmetry. For example, fivebranes can lie
along quaternionic planes in $R^8$ and preserve $(3,0)$ supersymmetry.
These are planes that are complex with respect to three complex structures
(eg the plane $-e^{1256}$ with respect to the hyper-K\"ahler structure
\p{hkstruct}). Supergravity solutions for these and other similar
configurations were discussed in \cite{ggpt,pt,pt2}.

\section{Acknowledgements}
We thank Gary Gibbons and especially Daniel Waldram for discussions.
JPG thanks the EPSRC for partial support.
Both authors are supported in part by PPARC through SPG $\#$613.    

%%%%%%%%%%%%%%%%%%%%

\end{document}